\documentclass[12pt]{jpconf}
\usepackage{iopams}

\def\Journal#1#2#3#4{{#4} {\it #1} {\bf #2}, #3 }
\def\a{\alpha}
\def\b{\beta}

\def\d{\delta}
\def\e{\epsilon}
\def\diver{\textrm{div}}
\def\curl{\textrm{curl}}

\begin{document}

\title{Purely radiative irrotational dust spacetimes}

\author{N Van den Bergh, B Bastiaensen, H R Karimian and L Wylleman}

\address{Faculty of Applied Sciences TW16, Gent University, Galglaan 2, 9000 Gent, Belgium}

\begin{abstract}
We consider irrotational dust spacetimes in the full non-linear regime which are "purely radiative" in the sense that the gravitational field satisfies the covariant transverse conditions $\diver(H)=\diver(E)=0$. Within this family we show that the Bianchi class A spatially homogeneous dust models are uniquely characterised by the condition that $H$ is diagonal in the shear-eigenframe.
\end{abstract}

\section{Introduction}

Irrotational dust spacetimes are typically considered as models for
the late universe or for gravitational collapse. They are
covariantly characterized by the dust four-velocity $u^a$, energy
density $\rho$, expansion $\theta$, and shear $\sigma_{ab}$, and by
the free gravitational field, described by the electric and magnetic
parts of the Weyl tensor $C_{abcd}$:
\begin{equation}\label{EenH}
E_{ab}\equiv C_{acbd}u^c u^d=E_{<ab>}   \quad  H_{ab}\equiv \frac{1}{2}{\varepsilon_{acd}}{C^{cd}}_{be}u^e=H_{<ab>}
\end{equation}
where $\varepsilon_{abc}=\eta_{abcd}u^d$ is the spatial projection of the spacetime permutation tensor $\eta_{abcd}$ and $S_{<ab>}={h_a}^c{h_b}^dS_{(cd)}-\frac{1}{3}S_{cd}h^{cd}h_{ab}$ is the projected, symmetric and trace-free part of $S_{ab}$ with $h_{ab}=g_{ab}+u_au_b$ the spatial projector into the comoving rest space, and $g_{ab}$ the spacetime metric. Gravitational radiation is covariantly described by the nonlocal fields $E_{ab}$, the tidal part of the curvature which generalizes the Newtonian tidal tensor, and by $H_{ab}$, which has no Newtonian analogue~\cite{Ellis}. As such, $H_{ab}$ may be considered as the true gravity wave tensor, since there is no gravitational radiation in Newtonian theory. However at least in the linear regime, as in electromagnetic theory, gravity waves are characterized by $H_{ab}$ and $E_{ab}$, where both are divergence-free but neither is curl-free \cite{Hogan}. Here the covariant spatial divergence and curl for tensors are defined by \cite{Maartens2}:

\[(\diver S)_a = D^b S_{ab}, \quad \curl S_{ab}=\varepsilon_{cd(a}D^c{S_{b)}}^d,
\]
where the spatial part $D_a$ of the covariant derivative $\nabla_a$ is given by
\[D_aA_{b\cdots}={h_a}^c{h_b}^d\cdots\nabla_cA_{d\cdots}
\]

For this reason spacetimes in which $\diver(E)=0=\diver(H)$ have been termed \emph{purely radiative}~\cite{Sopuerta}. Although, when considering the gravitational dynamics of matter in the full non-linear regime, these conditions give rise to two independent chains of rather severe integrability conditions, it was remarked in \cite{Sopuerta} that both are satisfied by the Bianchi class A solutions and hence purely radiative solutions at least can exist. We turn the argument around and show that, within the family of irrotational dust spacetimes for which $\diver(E)=0=\diver(H)$, the Bianchi class A spatially homogeneous dust models are uniquely characterised by the condition that H is diagonal in the common shear and E-eigenframe. The existence of the latter eigenframe is guaranteed by the Bianchi identity (\ref{Bianchi3}) below,
\[
 (\diver{H})_a=0\Longleftrightarrow[{E},\sigma]_a=0,
\]
where the spatial dual of the commutator of tensors, $[S,T]_a$, is
defined as $[S,T]_a=\varepsilon_{abc}S^{bd}{T_d}^c$.

\section{Basic variables and equations}

We follow the notations and conventions of the
Ellis-MacCallum orthonormal tetrad formalism~\cite{MacCallum}.
Herein, the normalized 4-velocity $u^a$ plays the role of the
timelike basis vectorfield $e_0^a$ of the tetrad
$\{e_0^a,\mathbf{e}_\alpha^a\}$. Greek indices take the values 1,
2 and 3 and refer to tetrad components with respect to
$\mathbf{e}^a_{\alpha}$.

The basic variables in this formalism are 24 independent linear
combinations of the Ricci rotation coefficients or, equivalently, of
the commutator coefficients $\gamma^a_{bd}$ associated with the
tetrad; these are the objects $n_{\a\b}$ and $a_\a$ defined by
$\gamma^\a_{\b\d}=\epsilon_{\b\d\e}n^{\e\a}+\delta^\a_\d a_\b -
\delta^\a_\b a_\d$, the components $\Omega_\a$ of the angular
velocity of the triad $\mathbf{e}_\a$ with respect to the local
`inertial compass' and the components $\dot{u}_\a$,
$\omega_{\alpha}$ and $\theta_{\a\b}$ of the spatial kinematic
quantities (acceleration vector, vorticity vector and expansion
tensor respectively). We also use the decomposition
$\theta_{\a\b}=\sigma_{\a\b}+1/3 \theta\delta_{\a\b}$, with
$\sigma_{ab}$ the trace-free shear tensor and $\theta$ the expansion
scalar. It turns out that there is a slight notational advantage in
using instead of the variables $n_{\alpha \beta} $ and $a_\alpha$
the connection coefficients $n_\alpha$, $q_\alpha$ and $r_\alpha$
related to the former by
\begin{equation*}
n_{\alpha +1\ \alpha-1}=(r_\alpha+q_\alpha)/2,\quad
a_\alpha=(r_\alpha-q_\alpha)/2,\quad n_{\alpha
\alpha}=n_{\alpha+1}+n_{\alpha-1}
\end{equation*}
(these expressions have to be read modulo 3, so for
example $\alpha = 3$ gives $n_{12}=2 q_3 + r_3$). For sake of simplicity we define $Z_{\alpha}$ and $m_\alpha$ as the spatial gradients of $\theta$ and $\rho$ respectively: $Z_{\alpha}= \partial_{\alpha}\theta$, $m_\alpha=\partial_{\alpha}\rho$.

Together with the Jacobi identities and the $(\alpha\beta)$ field
equations \cite{MacCallum}, the basic equations in the formalism are
the Ricci equations for $u_a$ and the Bianchi equations\cite{Ellis},
which for $\omega_{\alpha}=\dot{u}_{\alpha}=0$ simplify to the
following:

\subsection*{Ricci equations}
 \begin{equation}\label{Ricci1}
 \diver\sigma_a-\frac{2}{3}D_a\theta=0,
 \end{equation}
 \begin{equation}\label{Ricci2}
 \curl\sigma_{ab}-H_{ab}=0,
 \end{equation}
 \begin{equation}\label{Ricci3}
 \dot\sigma_{<ab>}+\frac{2}{3}\theta\sigma_{ab}+\sigma_{c<a}{\sigma_{b>}}^c+E_{ab}=0,
 \end{equation}
 \begin{equation}\label{Ricci4}
 \dot\theta+\frac{1}{3}\theta^2+\sigma_{ab}\sigma^{ab}+\frac{1}{2}(\rho+3p)=0,
 \end{equation}

\subsection*{Bianchi equations}
 \begin{equation}\label{Bianchi1}
 \dot\rho=-(\rho+p)\theta,
 \end{equation}
 \begin{equation}\label{Bianchi2}
 \diver E_a=[\sigma,H]_a+\frac{1}{3}D_a\rho,
 \end{equation}
 \begin{equation}\label{Bianchi3}
 \diver H_a=-[\sigma,E]_a,
 \end{equation}
 \begin{equation}\label{Bianchi4}
 \dot E_{<ab>}-\curl H_{ab}=-\theta E_{ab}+3\sigma_{c<a}{E_{b>}}^c-\frac{1}{2}(\rho+p)\sigma_{ab},
 \end{equation}
 \begin{equation}\label{Bianchi5}
 \dot H_{<ab>}-\curl E_{ab}=-\theta H_{ab}+3\sigma_{c<a}{H_{b>}}^c .
 \end{equation}

As we are working in the $(\sigma,E)$-eigenframe we derive from (\ref{Ricci3}) with $\alpha\neq\beta$ that,
\begin{equation}\label{angular}\left.\begin{array}{lll}
\Omega_1(\sigma_{22}-\sigma_{33})~=\Omega_2(\sigma_{33}-\sigma_{11})~=\Omega_3(\sigma_{11}-\sigma_{22})=0.\\
\end{array}\right.
\end{equation}
As a consequence the angular velocity $\Omega_{\alpha}$ can be assumed
to be 0: either all components automatically vanish, or, when e.g.~the $(e_2,e_3)$ plane is a shear
eigen-plane, then $\Omega_2=\Omega_3=0$ and we can choose an extra
rotation in this plane to make $\Omega_1=0$. The frame is then fixed
up to rotations in the $(e_2,e_3)$-plane for which the rotation
angle $\varphi$ satisfies $\partial_0\varphi=0$.

From the diagonal components of (\ref{Ricci2}) we obtain the following algebraic relations between $H_{\alpha\alpha}$,$\sigma_{\alpha\alpha}$ and $n_{\alpha}$:
\begin{equation}\label{Ricci21}\left.\begin{array}{lll}
H_{11}&=& n_3 (\sigma_{22}-\sigma_{11})+n_2 (\sigma_{33}-\sigma_{11}), \\
H_{22}&=& n_1 (\sigma_{33}-\sigma_{22})+n_3 (\sigma_{11}-\sigma_{22}), \\
H_{33}&=& n_2 (\sigma_{11}-\sigma_{33})+n_1 (\sigma_{22}-\sigma_{33}). \\
\end{array}\right.
\end{equation}
From the definition of $\diver E_{\alpha}$ and the assumption of $\diver(E)=0$ we have
\begin{equation}\label{defdiv} \left. \begin{array}{lll}
\partial_1E_{11}&=& E_{11}(r_1-2q_1)-E_{22}(r_1+q_1), \\
\partial_2E_{22}&=& E_{22}(r_2-2q_2)-E_{33}(r_2+q_2), \\
\partial_3E_{33}&=& E_{33}(r_3-2q_3)-E_{11}(r_3+q_3), \\
\end{array} \right.
\end{equation}
while the Bianchi equation (\ref{Bianchi2}) yields
\begin{equation}\label{Bianchi22} \left. \begin{array}{lll}
m_1&=&3H_{23}(\sigma_{33}-\sigma_{22}),\\
m_2&=&3H_{13}(\sigma_{11}-\sigma_{33}),\\
m_3&=&3H_{12}(\sigma_{22}-\sigma_{11}).\\
\end{array} \right.
\end{equation}
From (\ref{Ricci1}) we can derive the spatial derivatives of
$\sigma_{\alpha\alpha}$
\begin{equation}\label{dsigma} \left. \begin{array}{lll}
\partial_1\sigma_{11}&=&\frac{2}{3}Z_1+r_1(\sigma_{11}-\sigma_{22})-q_1(\sigma_{11}-\sigma_{33}),\\
\partial_2\sigma_{22}&=&\frac{2}{3}Z_2+r_2(\sigma_{22}-\sigma_{33})-q_2(\sigma_{22}-\sigma_{11}),\\
\partial_3\sigma_{33}&=&\frac{2}{3}Z_3+r_3(\sigma_{33}-\sigma_{11})-q_3(\sigma_{33}-\sigma_{22}).
\end{array} \right.
\end{equation}
Note that we can assume that the shear is non-zero, as otherwise
solutions would be conformally flat and we have a FRW universe.

We now act with the commutator $[\partial_0,\partial_{\alpha}]$ on
$\rho$ to obtain the relation
\begin{equation}
\partial_0 m_\alpha = -m_\alpha(\sigma_{\alpha \alpha}+\frac{4}{3} \theta) -(\rho+p) Z_\alpha,
\end{equation}
after which the time evolution of (\ref{Bianchi22}) results in
\begin{equation}\label{hdot}
\begin{array}{lll}
3(\sigma_{11}-\sigma_{22})\partial_0 H_{12} &=& (6\sigma_{11}^2-6\sigma_{22}^2-6E_{22}-3E_{33}+2\sigma_{22}\theta-2\sigma_{11}\theta)H_{12}+(\rho+p)Z_3\\

3(\sigma_{22}-\sigma_{33})\partial_0 H_{23} &=& (6\sigma_{22}^2-6\sigma_{33}^2-6E_{33}-3E_{11}+2\sigma_{33}\theta-2\sigma_{22}\theta)H_{23}+(\rho+p)Z_1\\

3(\sigma_{33}-\sigma_{11})\partial_0 H_{13} &=&
(6\sigma_{33}^2-6\sigma_{11}^2-  6
E_{11}-3E_{22}+2\sigma_{11}\theta-2\sigma_{33}\theta)H_{13}+(\rho+p)Z_2
\end{array}
\end{equation}

Provided we assume that also $H_{\alpha\beta}$ is diagonal in the
$(\sigma, E)$-eigenframe, it follows directly from
(\ref{Bianchi22}) and (\ref{hdot}) that $Z_{\alpha}=0$ and
$\partial_{\alpha}\rho=0$, suggesting that the corresponding
spacetimes \emph{might} be spatially homogeneous. Furthermore, as
$[\partial_0,\partial_{\alpha}] \equiv
-(\sigma_{\alpha\alpha}+\frac{1}{3}\theta)\partial_{\alpha}$ we obtain from
(\ref{Ricci4}) and
$\partial_{\alpha}\theta=0$ that $\partial_\alpha \sigma_{ab}\sigma^{ab} =0$ and hence, using (\ref{Ricci2}) and (\ref{dsigma}),

\begin{equation}\label{rq1}
\begin{array}{lll}
r_1(\sigma_{11}-\sigma_{22})^2-q_1(\sigma_{11}-\sigma_{33})^2 &=& 0\\
r_2(\sigma_{22}-\sigma_{33})^2-q_2(\sigma_{22}-\sigma_{11})^2 &=& 0\\
r_3(\sigma_{11}-\sigma_{33})^2-q_3(\sigma_{33}-\sigma_{22})^2 &=& 0 .\\
\end{array}
\end{equation}

We will treat the two cases of non-degenerate shear and degenerate
shear now separately.

\section{Non degenerate shear}

Propagating the last three equations along the fluid flow we get
\begin{equation}\label{rq2}
q_1(\sigma_{11}-\sigma_{33})\chi~=q_2(\sigma_{11}-\sigma_{22})\chi~=q_3(\sigma_{22}-\sigma_{33})\chi=0\\
\end{equation}
where $\chi$ is defined as
\[\chi\equiv\frac{2}{3}[E_{11}(\sigma_{22}-\sigma_{33})+E_{22}(\sigma_{33}-\sigma_{11})+E_{33}(\sigma_{11}-\sigma_{22})]
-(\sigma_{11}-\sigma_{22})(\sigma_{22}-\sigma_{33})(\sigma_{33}-\sigma_{11}) .
\]If $q_1^2+q_2^2+q_3^2\neq 0$ it would follow from (\ref{rq2}) that $\chi=0$, after which $\partial_1\chi=0$ leads to the contradiction $q_1\sigma_{11}(\sigma_{11}-\sigma_{33})=0$. Therefore we conclude that $q_{\alpha}=0$ and, according to (\ref{rq1}), we find that $r_{\alpha}=0$, such that $a_{\alpha}=0$ and $n_{\alpha\beta}$ is diagonal. It is easy to check that all spatial derivatives of the rotation coefficients now vanish and, as the frame is invariantly defined, we obtain the non-degenerate spatially homogenous models of {\bf Bianchi class A} (i.e.~types $I, II,VI_0,VII_0,VIII,IX$).

\section{Degenerate shear}
From the degeneracy of the shear it is easily verified that $E_{\alpha\beta}$ must be degenerate too; so we can assume
\[\begin{array}{lll}\sigma_{\alpha\beta}&=&\textrm{diag}(-2\sigma,\sigma,\sigma)\\
  E_{\alpha\beta}&=&\textrm{diag}(-2E,E,E).
  \end{array}
\]

Assuming that $H_{\alpha\beta}$ is diagonal in the
$(\sigma,E)$-eigenframe, (\ref{rq1}) implies

\begin{equation}\label{Hcond}
r_1-q_1 = r_3 = q_2 = 0
\end{equation}
while the (23) component of (\ref{Ricci2}) shows that $r_1+q_1=0$
and hence $r_1=q_1=0$.

At this point the only variables having possibly non-vanishing
spatial derivatives are $q_3,r_2$ and $n_1,n_2,n_3$. Eliminating
$\dot E$ from(\ref{Bianchi4}) for $\alpha=\beta=1$ and
$\alpha=\beta=2$, we obtain the following equation
\[0=H_{11}(-2n_1+n_2-2n_3)+H_{33}(-4n_1-n_2-n_3),
\]In combination with (\ref{Ricci21}) the latter leads to
\begin{equation}\label{vwn}(n_{1}+n_{2}+n_{3})(n_{2}-n_{3})=0.
\end{equation}
We will discuss the three cases that follow from (\ref{vwn})
separately; i.e.~$n_2\neq n_3$, $n_2=n_3\neq 0$ and $n_2=n_3=0$.
Note that in the ensuing classification $E$ is nowhere
zero, due to the non-existence of so-called Anti-Newtonian
universes \cite{Lode}

\subsection{$n_2-n_3\neq 0$}
In this case $n_1+n_2+n_3=0$, according to (\ref{vwn}). Propagating along $\mathbf{e}_0$ gives
\[0=\partial_0(n_1+n_2+n_3) = -3(n_2+n_3)\sigma\]and thus $n_1=0$ and $n_2=-n_3\neq0$. Applying $[\partial_0,\partial_3]$ to $\sigma$, gives us $4 n_2^2q_3=0$ and hence $q_3=0$, the $\partial_1$ derivative of which results in $r_2=0$. So $r_{\alpha}$ and $q_{\alpha}$ all become zero and we have
\[\begin{array}{lll}
n_{\alpha\beta} &=& \textrm{diag}(0,-n_2,n_2) \quad (n_2\neq0)\\
a_{\alpha} &=& 0.
\end{array}
\]This implies the vanishing of the spatial derivatives of all rotation coefficients($\partial_{\alpha}\equiv0$) and we obtain a spatially homogeneous universe of {\bf Bianchi class A, type $VI_0$}.

\subsection{$n_2 = n_3\neq0$}
First notice that a rotation about an angle $\varphi$ exists such that $n_1=-n_2$, $q_3=0$ and $\partial_2r_2\equiv r_2^2+K=0$, where
\begin{equation}\label{K}K\equiv\rho+3\sigma^2-\frac{1}{3}\theta^2+n_2^2.\end{equation}
This requires the existence of a solution $\varphi$ of the system
\begin{equation}\label{rot1}\left\{\begin{array}{lll}
\partial_0\varphi &=& 0\\
\partial_1\varphi &=& n_1+n_2\\
\partial_2\varphi &=& \sin(\varphi)\sqrt{-K}-q_3\\
\partial_3\varphi &=& \cos(\varphi)\sqrt{-K}-r_2
\end{array}\right.
\end{equation}
the integrability conditions of which are identically satisfied (the first equation of (\ref{rot1}) guarantees that $\Omega_1$ remains zero).

Note that the system (\ref{rot1}) can only give us a solution when $K\leq 0$. In that case we obtain
\[n_{\alpha\beta}=\left[\begin{array}{ccc}2n_2 &0&\sqrt{-K}/2\\
0&0&0\\
\sqrt{-K}/2&0&0\end{array}\right], \quad a_{\alpha} = \left[\begin{array}{c}0\\ \sqrt{-K}/2\\0\end{array}\right]
\]
and we have a LRS spacetime with
\begin{enumerate}
    \item {\bf $K=0\Rightarrow r_2=0$: Bianchi class A, type $II$},
    \item {\bf $K<0\Rightarrow r_2\neq 0 $:} when we look at the Ellis-MacCallum quantity \cite{Ellis3} $T\equiv4(a_2^2+a_3^2)$, we find $T=r_2^2$ and hence this class is equivalent with {\bf Bianchi class A, type $VIII$}.
\end{enumerate}

When $K>0$ (\ref{rot1}) has no solution, but now we can look for another rotation $\varphi$ to make $a_{\alpha}=0$. This requires the existence of a solution $\varphi$ of the system
\begin{equation}\label{rot2}\left\{\begin{array}{lll}
\partial_0\varphi &=& 0\\
\partial_1\varphi &=& -\frac{K}{2n_2}+n_1+n_2\\
\partial_2\varphi &=& -q_3\\
\partial_3\varphi &=& -r_2
\end{array}\right.
\end{equation}
Again the integrability conditions are identically satisfied and we find
\[a_{\alpha}=0, \quad n_{\alpha\beta}=\textrm{diag}(2n_2,\frac{K}{2n_2},\frac{K}{2n_2}),
\]
leading to the LRS {\bf Bianchi class A, type $IX$} models.

\subsection{$n_2=n_3=0$}
These models are exceptional in the sense that they are `purely electric' $(H_{ab}=0)$.

As in (4.2) one shows that a rotation exists under which $q_3$, $n_1$ and $\partial_2 r_2 =r_2^2+K$ become zero,
where $K$ is the same quantity as defined by (\ref{K}).

When $K<0$, we now obtain
\[n_{\alpha\beta}=\left[\begin{array}{ccc} 0 &0&\sqrt{-K}/2\\
0&0&0\\
\sqrt{-K}/2&0&0\end{array}\right], \quad a_{\alpha} = \left[\begin{array}{c}0\\ \sqrt{-K}/2\\0\end{array}\right]
\]

which is a LRS spatially homogenous {\bf Bianchi class B, type $VI_{-1}$} model and which is pseudo-spherically symmetric \cite{Ellis3}.

When $K=0$, then $r_2=0$ and we have $a_{\alpha}=0$, $n_{\alpha\beta}=0$, which is a LRS spatially homogenous {\bf Bianchi class A, type $I$} model (equivalent to {\bf Bianchi class A, type $VII_0$} \cite{Ellis3}).

A problem arises when $K>0$: $q_3$, $n_1$ and $n_2$ can still be made zero by a rotation, but it is no longer possible to choose a frame in which all spatial derivatives identically vanish. However the Cartan equations can now be integrated immediately and one finds the metric
\[ds^2=-dt^2+Q(t)^2dr^2+P(t)^2[d\theta^2+\sin^2(\theta) d\phi^2].\]
The resulting models are the Kantowski-Sachs universes, which, as is well known, do not admit a $3$-dimensional isometry group acting simply transitively on the $t=\textrm{constant}$ hypersurfaces.

\section{Conclusion}
We have shown that the Bianchi class A dust models can be uniquely characterised as irrotational dust spacetimes which are purely radiative in the sense that the gravitational field satisfies $\diver{(H)}=\diver{(E)}=0$ under the assumption that also the magnetic part of the Weyl tensor $H_{\alpha\beta}$ is diagonal in the shear-electric eigenframe (and hence $[H,E]=0$). The only possible exception arises for degenerate shear and $n_{\alpha\alpha}=0$: when $K\equiv \rho + 3\sigma^2-\frac{1}{3}\theta^2>0$ or $K<0$, in which case solutions are given respectively by the Kantowski-Sachs universes, or by the pseudo-spherically symmetric solutions of Bianchi class $VI_{-1}$. These models however are `purely electric' and cannot be called `purely radiative' in a strict sense.

The whole analysis presented here can be easily generalized to give a characterisation of spatially homogenous non-rotating \emph{perfect fluids} of Bianchi class A (to be submitted). The next logical step is to investigate the case where $[H,E]\neq 0$: it is hoped that new classes of cosmologically interesting solutions will turn up. Alternatively one would obtain the remarkable result that the Bianchi class A spacetimes  are the unique purely radiative ones (work still in progress).

\end{document}